\documentclass[usegraphicx,usedcolumn,usenatbib]{mn2e}
\usepackage{rotating}

\newif\ifAMStwofonts
\AMStwofontstrue

\ifCUPmtlplainloaded \else
  \ifAMStwofonts \else 

  \fi
\fi

\title[ISO-LWS two colour diagram of YSOs]
  {ISO-LWS two colour diagram of young stellar objects\thanks{ISO is an
ESA project with instruments funded by ESA Member States (especially the PI
countries: France, Germany, the Netherlands and the United Kingdom) and
with the participation of ISAS and NASA.}}
\author[S. Pezzuto et al.]
  {S.~Pezzuto,$^1$ F.~Grillo,$^1$ M.~Benedettini,$^1$ E.~Caux,$^2$ 
  A.M.~Di Giorgio,$^1$
  \newauthor T.~Giannini,$^3$ S.J.~Leeks,$^{4,5}$
  D.~Lorenzetti,$^3$ B.~Nisini,$^3$ P.~Saraceno,$^1$
  \newauthor  L.~Spinoglio,$^1$ E.~Tommasi$^6$\\
  $^1$Istituto di Fisica dello Spazio Interplanetario, CNR, Via Fosso del
      Cavaliere 100, 00133 Roma, Italy\\
  $^2$CESR, B.P. 4346, 31028 Toulouse Cedex 04, France\\
  $^3$Osservatorio di Roma, Via Frascati 33, 00040 Monte Porzio, Italy\\
  $^4$Queen Mary \& Westfield College, University of London, Mile End Road,
      London E1 4NS, UK\\
  $^5$Dept of Physics \& Astronomy, Cardiff University, PO Box 913, Cardiff,
  CF24 3YB, UK\\
  $^6$Agenzia Spaziale Italiana, V.le Liegi 26, 00198 Roma, Italy}
\date{Accepted . Received}
\pagerange{\pageref{firstpage}--\pageref{lastpage}}
\pubyear{}

\def\LaTeX{L\kern-.36em\raise.3ex\hbox{a}\kern-.15em
    T\kern-.1667em\lower.7ex\hbox{E}\kern-.125emX}

\begin{document}

\label{firstpage}

\maketitle

\begin{abstract}
We present a [60--100] vs. [100--170] $\mu$m two colour diagram for a sample of
61 young stellar objects (YSOs) observed with the Long Wavelength Spectrometer
(LWS) on-board the Infrared Space Observatory (ISO). The sample consists of 17
Class 0 sources, 15 Class~I, 9 Bright Class~I ($L_{\rm bol}>10^4\,L_\odot$), 20
Class~II (14 Herbig Ae/Be stars and 6 T Tauri stars). We find that each class
occupies a well defined region in our diagram with colour temperatures
increasing from Class~0 to Class~II. Therefore the [60--100] vs. [100--170] two
colours diagram is a powerful and simple tool to derive from future (e.g. with
the Herschel Space Observatory) photometric surveys the evolutionary status of
YSOs. The advantage over other tools already developed is that photometry at
other wavelengths is not required: three flux measurements are enough to derive
the evolutionary status of a source. As an example we use the colours of the YSO
IRAS 18148$-$0440 to classify it as Class~I. The main limitation of this work is
the low spatial resolution of the LWS which, for some objects, causes a high
uncertainty in the measured fluxes due to the background emission or to the
source confusion inside the LWS beam.
\end{abstract}


\section{Introduction}
It is generally accepted that the evolutionary status of young stellar objects
(YSOs) is characterised by the shape of their spectral energy distribution
(SED). During the early phases of star formation the emitted SED has the shape
of a cold, few tens of K, blackbody modified by the emissivity of the dust. The
central object, even when already formed, is not visible at any wavelength since
the surrounding dust envelope completely obscures the inner region. As the
evolution goes on, the optical depth of the dust decreases making the central
object optically visible until its spectrum looks like that of a normal star
with an infrared excess due to the residual circumstellar material. It is then
clear that the characterisation of the evolutionary status of YSOs is better
done by observing the dust emission rather than the stellar radiation which is
only visible during the late phases of star formation.

ISO, the Infrared Space Observatory \citep{kess}, has opened a new observing
window covering the near to far infrared wavelengths up to $\lambda\sim 200
\mu$m, where the dust emission peaks. It is then natural to use ISO data to
derive the evolutionary status of YSOs. To this aim we introduce a two colour
diagram defined between 60 and 170$\,\mu$m which exploits almost all the
spectral coverage of the Long Wavelength Spectrometer, LWS \citep{clegg}.
Already with IRAS data two colour diagrams have been used to infer the
evolutionary phase of YSOs, (e.g. \citealt{beich}, or \citealt{berri}), however,
the LWS extends the FIR (Far InfraRed) spectral coverage of IRAS by a factor
$\sim 2$ in the region where the peaks of emission of blackbodies with
temperatures in the range $15-30\,$K fall, typical of YSOs in the early phases
of their evolution.

\citet{vilspa} and \citet{parigi} have indeed already shown that such a diagram
allows to identify the evolutionary status of YSOs. While in the previous work a
small sample of objects was used, we have now completed the analysis of the
whole sample of YSOs observed by the LWS as part of the core program dedicated
to the study of the star formation. 

With respect to previous evolutionary diagnostic tools, like the spectral index
$\alpha$ \citep*{ALS}, the bolometric temperature (\citealt{ml},
\citealt{chen}), and the $L_{\rm bol}$ vs. $F_{\rm 1.3mm}$ diagram
\citep{paolo}, our method can be applied to YSOs in all evolutionary phases
(Class~0, I and II objects), and does not require the knowledge of either $L_{
\rm bol}$, or the complete SED.

In Section~2 we present the sample, and the data analysis is briefly reported;
in Section~3 the two colour diagram is shown and discussed; the conclusions are
summarised in Section~4.

\section{Observations and data reduction}
Our sample consists of 61 YSOs for which an LWS full range grating spectrum, $43
-197\,\mu$m with a resolution $\lambda/\Delta\lambda\sim 200$, was obtained. It
corresponds to almost all the observations carried out with the LWS as part of
the ISO-LWS core programme on star formation, excluding a few sources whose
spectra are too noisy to derive reliable photometric data. We have decided to
include for this paper only the sample of our guaranteed time, to have a
homogeneous set of data taken with the same observational procedure and
comparable integration times. Our original sample, to which the 61 objects
discussed here belong, was selected on the basis of the knowledge of the instrument (spatial and spectral resolution, sensitivity) and on the spectral characteristics of the known sources, at the time of the definition of the observing programmes (around 1995).

The raw data have been processed with version~8 of the off-line
pipeline\footnote{After submitting the manuscript, the final version of this
software has been released (version 10). The changes in the data processing with
respect to version 8 are not relevant to the present work, as we verified for
some faint or noisy sources (for instance SR~9).} and analysed with the ISAP
package. After removing bad points, the scans were averaged. The strong source correction, which takes into account the non-linearity of the detectors at high fluxes (\citealt{leeks}, \citealt{leeksthesis}), was applied to the data if it was necessary.

The LWS spectra are affected by instrumental effects (fringes, memory effects,
bad flat field and dark current removal, see \citealt{cecil}) which might potentially influence the shape of the spectrum. Such effects can be, at least
partially, corrected by using specific routines (LIA, the LWS Interactive
Analysis). However it has been verified \citep{franz} that for the majority of
our objects, the fluxes are not considerably changed by applying these routines.
In the end, we adopted the fluxes computed without any correction.

For some objects, one or more off-source positions have been observed to
estimate the background emission. In our previous work \citep{parigi} we
compared the colours before and after removing the background and we found that,
again in a statistical sense, the subtraction does not alter our results.
Moreover, in star forming regions the presence of a diffuse, not homogeneous,
cloud emission makes it difficult to define the background level. In addition,
when the instrument does not have enough spatial resolution, the confusion due
to other sources in the beam can make even the meaning of background dubious.
For these reasons we did not subtract the background from the on-source spectra.
The large spread of the colours in our diagram is, likely, the consequence of
this choice, as we discuss in the next section.

The colour $[\lambda_i-\lambda_j]$ is defined as:
\begin{displaymath}
[\lambda_i -\lambda_j]=\log\frac{F_j}{F_i}
\end{displaymath}
$F_{i,j}$ being the flux densities in Jy, averaged over a certain band, at the
wavelengths $\lambda_{i,j}$. In the case of photometric data $\lambda$ is the
effective wavelength of the adopted band. Dealing with spectra the choice of the
wavelength is somewhat arbitrary. A natural choice is to measure the flux at $60
\,\mu$m and $100\,\mu$m, which allows a comparison with IRAS data. As a third
wavelength we adopted $170\,\mu$m: this choice was first motivated by the
instrumental characteristics of the LWS (as discussed in \citealt{vilspa}). It
then turned out that these three wavelengths are close to the effective
wavelengths of the broadband filters of PACS \citep{ap}, a far infrared camera
for imaging and spectroscopy which is part of the payload of the future ESA
cornerstone mission the Herschel Space Observatory. In such a way it will be
possible to use our two colour diagram with PACS data.

For the bandwidths there is no obvious choice. At first sight a good solution
could be to simulate the spectral response of the IRAS bands, as the ISAP
package allows. There are, however, at least two reasons for not following this
approach. The first one is that this choice is not applicable to the third band
since IRAS did not observe at $170\,\mu$m. Secondly, future space missions will
enlarge the number of known YSOs, so it is important that our tool does not
depend on the photometric bands adopted in previous observations. So, we simply
averaged the spectra over 1$\,\mu$m around the three wavelengths, and verified
that our conclusion are not altered when data derived with a completely
different instrument, in particular with IRAS, are used. This topic will be
addressed in the next section.

The uncertainty of the LWS fluxes can be as low as 10\% for on-axis, not too
bright nor too faint, point sources. These conditions are seldom met for our
sources, so we adopted a more conservative error of 30\%, which turns into an
error on each colour of 0.26.

\begin{table*}
\centering
\hbox{
  \rotcaption{The sample of YSOs observed. Sources have been grouped according
  to their known evolutionary status. Inside each group sources have been sorted
  in increasing right ascension. An asterisk flags objects classified as FU
  Orionis. The [60-100] IRAS colour has been computed by using the fluxes
  reported in the SIMBAD database or the fluxes given in the quoted reference.
  The IRAS colour of L~1448~N has been obtained by adding the fluxes of L~1448~N
  and L~1448~NW, which fall inside the LWS beam.}
  \label{tab:table}
\begin{sideways}
    \begin{tabular}[b]{lc...lc...}
      \hline \\[-5pt]
&IRAS ID&\multicolumn{2}{c}{[60--100]}&\multicolumn{1}{c}{[100--170]}&
&IRAS ID&\multicolumn{2}{c}{[60--100]}&\multicolumn{1}{c}{[100--170]}\\[+2pt]
&&\multicolumn{1}{c}{LWS}&\multicolumn{1}{c}{IRAS}&&
&&\multicolumn{1}{c}{LWS}&\multicolumn{1}{c}{IRAS}\\
\bf Class 0      &&&&&\bf Bright Class~I\\[+2pt]
L 1448 N         &&0.53&0.54^1&0.20&NGC 281 W&00494+5617&0.39&0.38^8&0.025\\
L 1448 mm        &&0.53&&0.094&AFGL 437&03035+5819&0.0096&0.089&-0.19\\
NGC 1333 IRAS 2  &03258+3104&0.30&0.46^2&-0.077&AFGL 490&03236+5836&0.025&0.039&-0.12\\
NGC 1333 IRAS 6  &&0.50&0.52^2&0.38&NGC 2024 IRS2&&-0.11&&-0.23\\
NGC 1333 IRAS 4  &&0.84&0.68^2&0.22&NGC 6334 I   &&0.14&&-0.13\\
&03282+3035  &0.57&0.77&0.28&W 28 A2&&-0.0016&&-0.24\\
VLA 1            &&0.54&&0.24 &M 8 E&&0.19&&-0.11\\
NGC 2024 FIR 3   &&0.00&&-0.20&GGD 27 IRS&&-0.0084&&-0.20\\
NGC 2024 FIR 5   &&-0.078&&-0.21&S 87 IRS 1&19442+2428&0.025&0.18^9&-0.12\\
HH 25 mm         &&0.57&&0.19\\
HH 24 mm         &&0.47&&0.23&\bf Herbig Ae/Be\\
VLA 1623         &&0.52&&0.20  &LK H$\alpha$ 198&00087+5833&0.21&0.17^{10}&-0.037\\
&16293$-$2422&0.55&0.61&0.12 &V376 Cas&&0.067&0.16^{10}&-0.086\\
L483&18148$-$0440&0.39&0.27&0.063&*Z CMa&07013$-$1128&0.091&0.041&-0.091\\
Serpens FIRS 1&18273+0113	 &0.53&0.41^3&0.081&HD 97048&11066$-$7722&0.11&0.048&-0.014\\
B 335            &19345+0727&0.69&0.70&0.19 &DK Cha&12496$-$7650&-0.0026&-0.082&-0.032\\
L 723            &19156+1906&0.58&0.48&0.20 &CoD $-39^\circ$ 8581&13547$-$3944&0.54&0.62^{11}&0.10\\
                         &&&     &&CoD $-42^\circ$ 11721&16555$-$4237&0.087&0.062&-0.013\\
\bf Class~I      &&&             &&MWC 297&18250$-$0351&-0.062&0.30&-0.27\\
SVS 13           &03259+3105&0.26&0.39^2&0.028&R CrA&18585$-$3701&0.27&0.30&-0.060\\
$^*$L 1551 IRS 5 &04287+1801&0.15&0.089&-0.043&T CrA&&0.43&&0.26\\
L1641 N          &05338$-$0624&0.40&0.38&0.12 &PV Cep&20453+6746&0.12&0.064&0.0023\\
HH 26 IRS        &&0.28&&0.27  &V645 Cyg&21381+5000&0.098&0.089&-0.084\\
HH 46 IRS        &08242$-$5050&0.34&0.35&0.089&LK H$\alpha$ 234&21418+6552&0.16&0.25&-0.14\\
WL 16            &&0.23&&0.071&MWC 1080&23152+6034&0.21&0.46&-0.094\\
Elias 29         &&0.21&0.41^4&0.098\\
$\rho$ Oph IRS 43&&0.30&&0.26&\bf T Tauri\\
$\rho$ Oph IRS 44&&0.26&&0.24&*RNO 1B&&0.34&&0.016\\
Re 13	         &16289$-$4449&0.26&0.53^5&0.13&T Tau&04190+1924&0.071&-0.0026&-0.12\\
$^*$HH 57 IRS    &16289$-$4449&0.17&-0.054^5&0.071&DG Tau&04240+2559&0.071&0.070&0.059\\
L 379 IRS 3      &18265$-$1517&0.39&0.46^6&0.088&HL Tau&04287+1807&0.055&<0.77&-0.057\\
IC 1396 N        &&0.41&&0.11&SR 9&&0.53&&0.18\\
NGC 7129 FIRS 2  &&0.36&0.33^7&0.077&*Elias 1-12&21454+4718&0.40&0.32&0.18\\
L 1206           &22272+6358&0.21&0.28&-0.10\\[+2pt]
\hline
\multicolumn{10}{p{190mm}}{$^1$\citet{bar}; $^2$\citet{jenning} (actually
[50-100]); $^3$\citet{hb}; $^4$\citet{young}; $^5$\citet{tp};
$^6$\citet{hilton}; $^7$\citet{epc}; $^8$\citet{mook}; $^9$\citet{ckmg};
$^{10}$\citet{hsvk}; $^{11}$\citet{glen}}
      \end{tabular}
\end{sideways}
}
\end{table*}

\section{Results and discussion}
In Table~\ref{tab:table} we report, for each object: its name; the identifier in
the IRAS Point Source Catalogue, if present; the LWS and, if available, the IRAS
colour $[60-100]$; the LWS $[100-170]$ colour. The membership of each source to
a class has been assessed by looking at the literature.

\begin{figure*}
  \includegraphics[width=15cm]{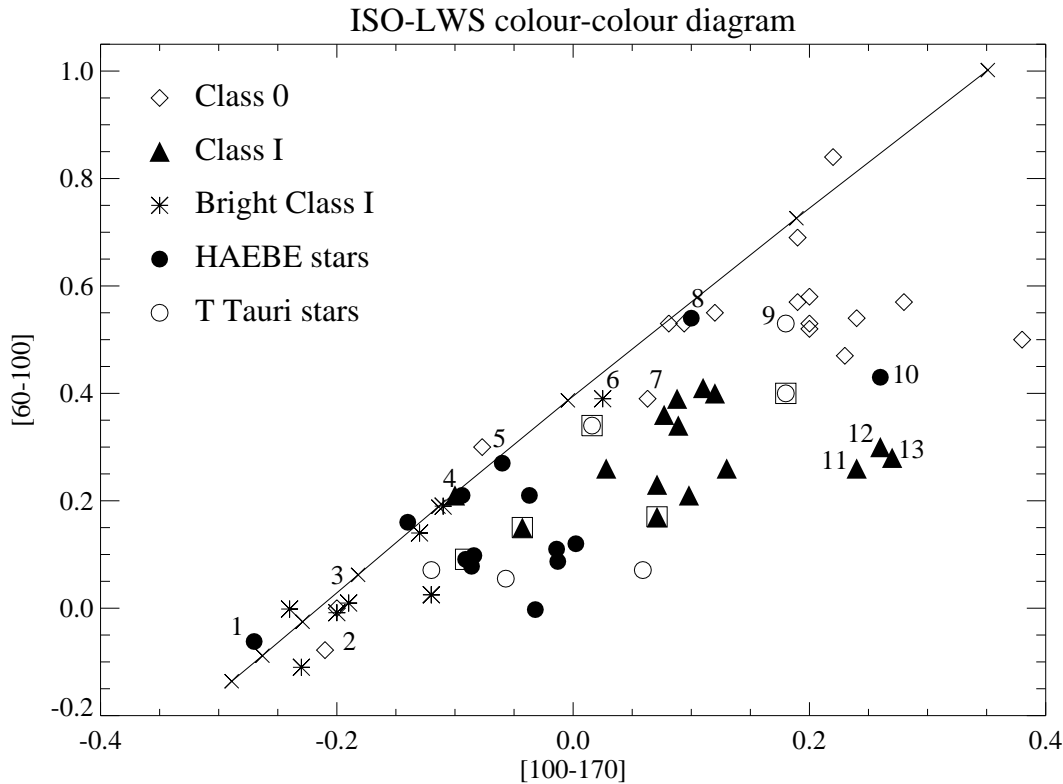}
  \caption{Two colour diagram for the sample of YSOs. Bright Class~I sources
   are those with bolometric luminosity $L_{\rm bol}>10^4\,L_\odot$. Boxed
   symbols identify FU Orionis objects. The error, not displayed, on each point
   is 0.26 in both colours, assuming an uncertainty on the fluxes of 30\%. Solid
   line: blackbody colours at temperatures (marked by crosses): 25\,K (upper
   right corner), 30, 40, 50, 60, 70, 80 and 90\,K (lower left corner). The
   numbers identify sources which are individually discussed in the text: 1)
   MWC 297; 2) FIR 5 and 3) FIR 3 (both in NGC 2024); 4) L 1206; 5) IRAS 2 (in
   NGC 1333); 6) NGC 281 W; 7) IRAS 18148-0440; 8) IRAS 13547-3944; 9) SR 9; 10)
   T CrA; 11) IRS 44 and 12) IRS 43 (both in $\rho$ Oph); 13) HH 26 IRS.}
  \label{fig1}
\end{figure*}

The [60--100] vs. [100--170] diagram is shown in Figure~\ref{fig1}. For
comparison we reported the colours, computed following the same procedure
outlined before, of a blackbody at various temperatures (marked by crosses and
connected by the solid line) ranging from 25~K (top right) to 90~K (bottom
left).

The FU Orionis sources are reported in the figure but not considered in the
following discussion since their colours are likely influenced by the outbursts
of their disc which represent transient phenomena in the process of star
formation.

The sources do not strictly follow the blackbody line, but there is a clear
trend of increasing colour temperature $T_{\rm c}$ with the age of the sources:
Class 0 objects have $T_{\rm c}\sim 30\,$K, Class I have $T_{\rm c}\sim 35-40
\,$K and for Class II $T_{\rm c}\sim 50\,$K. Bright Class I show the highest
temperatures with $T_{\rm c}\sim 50-80\,$K. This trend can be qualitatively
explained by considering that at these wavelengths $T_{\rm c}$ depends only on
the temperature stratification of the circumstellar matter. In the first stage
of the evolution we can see only the outer region of the envelope at cold
temperatures. The optical depth of the envelope decreases with the age so that
we can look deeper into the inner, and warmer, regions of the envelope and,
consequently, $T_{\rm c}$ increases.

The remaining of this section is organised as follows: first we discuss
separately each class and argue that the presence of outlyers, evident in the
diagram, can be tentatively explained, in many cases, as due to background
contamination. For such reason we cannot average all the colours for each class
because this procedure assumes that the dispersion of the points is caused only
by random fluctuactions. Instead, we preferred to compute the median which is
less sensitive than the mean to the outlyers. Then, in Section~3.4, we test our
result with a different data set, using IRAS fluxes, to verify that we did not
introduce any bias in the data analysis, at least for the $[60-100]$ colour.
Finally, the medians are reported in Figure~\ref{fig2} along with their root
mean square (RMS) errors, and their position with respect to the blackbody
colours is discussed in Section~3.5. 

\subsection{Class 0 sources}
Class 0 sources appear well clustered in the upper right region of the diagram
with $[60-100]\ga 0.45$ and $[100-170]\ga 0.08$, except four objects discussed
below. The median colours are: $[60-100]=0.53\pm 0.23$ and $[100-170]=0.19\pm
0.18$.

IRAS~18148--0440 (number 7 in Figure~\ref{fig1}) has a smaller $[100-170]$
colour and falls closer to Class~I sources. It is reported as a Class~0 in the
compilation by \citet{AWB2} but \citet{taf} argue that this source is already in
a transition state between Class~0 and I. From our diagram we suggest that
IRAS~18148 would be better classified as Class~I.

The two sources FIR~3 and FIR~5 in NGC~2024 (numbers 2 and 3 in the figure) have
blue colours, similar to those of Bright Class~I ($L_{\rm bol}> 10^4\,L_\odot$).
From the analysis of the same spectra used in our work, \citet{teresa} have
shown that in NGC~2024 the emission lines are mainly produced by the extended
cloud in which the sources are embedded. Since the H{\sc ii} region associated
with NGC~2024 has a luminosity of $2.8\cdot 10^4\,L_\odot$ \citep{merz}, while
for FIR~5 \citet{wiese} estimate $L_{\rm bol}\sim 1-4\,L_\odot$, from the
position of the sources in our diagram we suggest that the observed continua are
dominated by the diffuse PDR more than by the objects themselves.

The source IRAS 2 in NGC 1333 (labelled 5 in the plot) falls close to Class~II
objects. The IRAS colour found by \citet{jenning} and reported in
Table~\ref{tab:table} is redder than ours and puts the source closer to other
Class~0. The fluxes were measured with the Chopped Photometric Channel of IRAS
which provided a higher angular resolution than that specific to the IRAS
survey, but it is worth noting that the colour directly measured by IRAS is
0.35, not very different from our value of 0.30. It is then likely that the
colour we found is strongly affected by the low spatial resolution of the LWS
rather than depending on intrinsic characteristics of the source.

\subsection{Class~I and Bright Class~I sources}
Without considering the two FU Ori sources, we find for Class~I objects the
following median colours: $[60-100]=0.280\pm 0.075$ and $[100-170]=0.09\pm
0.10$. The large scatter in the $[100-170]$ colour is due to the position of
L~1206 (source number 4) close to the Class~II objects, and to a little
``cluster'' of three sources with $[100-170]>0.2$:  for IRS 43 and 44 (numbers
12 and 11 in the figure) the large $[100-170]$ colour is due to the diffuse
emission in the $\rho$~Oph complex in which they are embedded. In fact, the
subtraction of the background moves their colours into the Class~I region.

Objects having a total luminosity $L\ge 10^4\,L_\odot$ have been considered as
a distinct group (Bright Class~I), since such a high luminosity is usually
associated with very massive hot young stars whose FIR colours are expected to
be different from those of lower luminosity Class~I sources. In fact, with the
exception of NGC~281W (number 6), they have $[60-170]<0.2$ and $[100-170]<-0.1$,
and fall close to the blackbody line with colour temperatures $T\ge 50\,$K. The
median colours are $[60-100]=0.02\pm 0.16$ and $[100-170]=-0.130\pm 0.083$.

The colours of NGC~281W, whose total luminosity is $2.4\cdot 10^4\,L_\odot$
\citep*{carp}, are the reddest of this group and are in strong disagreement with
those of the other objects. We do not have any explanation for such a position.

\subsection{Class~II sources}
This group is divided into low mass stars, $M\la 2\,M_\odot$, called T~Tauri
after their prototype, and intermediate mass stars, $2\,M_\odot\la M\la 8\,
M_\odot$, called Herbig Ae/Be stars (or HAEBE). This division reflects physical
and observational differences between them, but in our diagram they occupy the
same region so, to increase the statistical significance of our work, we will
discuss them together. All the Class~II objects, with the exception of MWC~297,
IRAS 13547, T~CrA and SR 9, cluster in a well defined region of the plot. The
median colours of all sources, except the FU Ori, are $[60-100]=0.11\pm 0.14$
and $[100-170]=-0.03\pm 0.12$, with a colour temperature $T_{\rm c}\sim 50\,$K.

For SR~9 (number 10 in the figure) the fluxes in the on-source position are
equal to those in two observed off-source positions so that we conclude that the
continuum emission of this T~Tauri star is completely dominated by the intense
diffuse background of the $\rho$~Oph region, in which SR~9 is embedded, with
colours in agreement with those of Class~0 sources (see Table~\ref{tab:table}).
Its $[100-170]$ colour is similar to that of IRS 43 and 44 which also lie in
the same region (see the previous discussion).

The anomalous position of the HAEBE star MWC~297, in the bottom left corner and
labelled with 1, is compatible with the finding of \citet{drew} that it could be
in a more evolved stage and already on the main sequence. In the opposite
corner, IRAS~13547 (number 8) and T~CrA (10) have colours compatible with
Class~0 sources: even if their membership to HAEBE is doubtful, the former being
defined as ``potential candidate'' and the latter reported among ``other early
type emission line stars with IR excess'' in the compilation of HAEBE by
\citet*{the}, these sources are optically visible and they cannot belong to the
group of Class~0. The only conclusion that we can draw from our diagram is that
the FIR emission is either produced by a diffuse background, for T~CrA see the
FIR maps of \citet{wilk}, or by a close cold companion.

\subsection{Comparison with IRAS data}
As we already pointed out, it is important that the result of our work, i.e.
that the $[60-100]$ vs. $[100-170]$ diagram is able to separate YSOs belonging
to different evolutionary stages, is independent of the instrument used and on
the adopted photometric bands. Further, since we neglected instrumental effects,
did not take into account the background contamination, defined two colours
chosing three arbitrary wavelengths and averaging the fluxes over three
arbitrary bandwidths, the reader could wonder about the reproducibility of our
finding with different instrument/sample/data analysis.

To address some of these points we compared the median of the $[60-100]$ colours
we found, with the medians derived using the IRAS fluxes, when available in
literature. These two sets of data have in common only the effective
wavelengths: bandwidths and data acquisition/reduction are completely different.
As in our original sample, we did not consider FU Orionis objects. The two sets
of medians are reported in Table~\ref{comp}.

\begin{table}
  \caption{Comparison between median $[60-100]$ colours derived with LWS and                IRAS data.}
  \label{comp}
    \begin{tabular}{l.@{$\pm$}l.@{$\pm$}l}
      \hline\\
&\multicolumn{2}{c}{LWS}&\multicolumn{2}{c}{IRAS}\\ \hline
Class 0       &0.53 &0.23 &0.53 &0.16\\
Class~I       &0.280&0.075&0.385&0.078\\
Bright Class~I&0.02 &0.16 &0.13 &0.16\\
Class~II      &0.11 &0.14 &0.12 &0.20\\ \hline
\end{tabular}
\end{table}

For Bright Class~I the number of sources with IRAS data is too small to draw any
conclusion. For Class 0 and II the agreement between the two sets of data is
excellent. Only for Class~I is the difference larger than 1 RMS. From our point
of view, however, the difference between 0.280 and 0.385 does not make a big
difference in the conclusion that Class~I sources occupy a specific region of
the diagram. What this difference means is just that we cannot specify with
great accuracy what the median $[60-100]$ colour is.

\begin{figure*}
  \includegraphics[width=15cm]{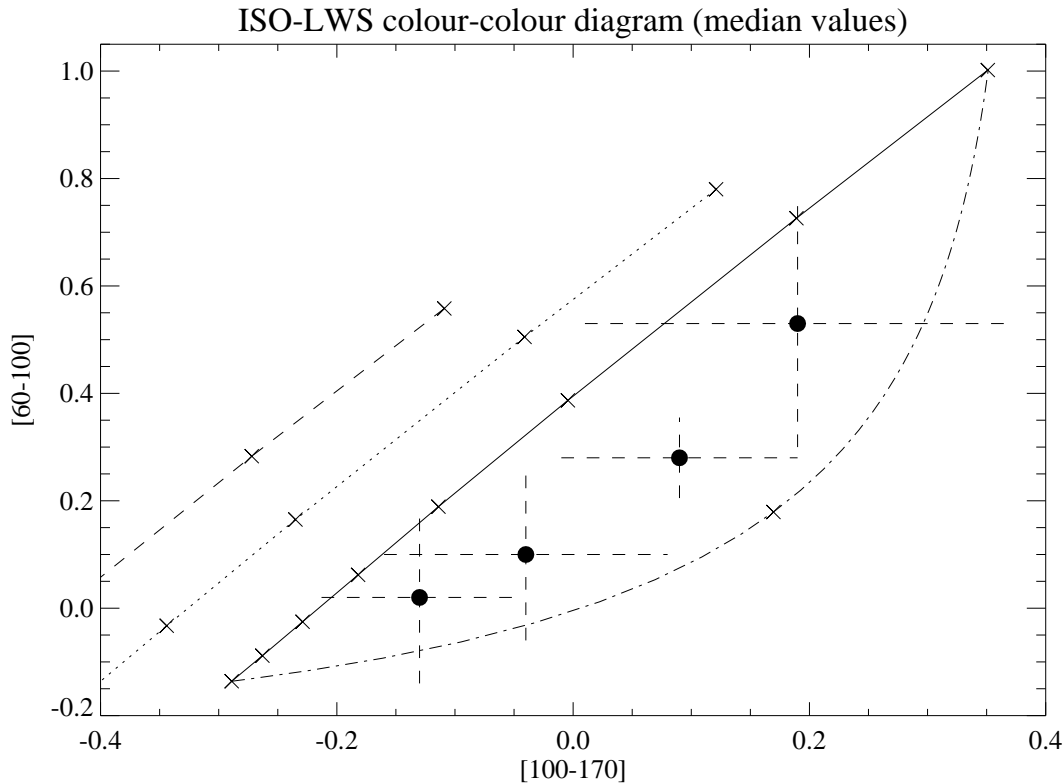}
  \caption{The median colours for each class (filled symbols) with their root
  mean square errors (dashed lines). From top to bottom: Class~0, Class~I,
  Class~II and Bright Class~I. Solid line: colours of a single blackbody as in
  Figure~1; dotted line: colours of a blackbody modified with an emissivity law
  $\propto\lambda^{-1}$ at $T$: 25, 30, 40, 50\,K; dashed line: as before but
  with the emissivity $\propto\lambda^{-2}$ and $T$: 25, 30\,K. Dotted-dashed
  line: colours of two blackbodies at 25 and 90\,K and different radii; the
  cross marks the colours corresponding to the case when the radius of the
  colder source is 10 times larger than the radius of the warmer source.}
  \label{fig2}
\end{figure*}

\subsection{The median colours: dust emissivity vs. source confusion}
The median colours are reported in Figure~\ref{fig2}. In this figure we also
plotted, as in Figure~\ref{fig1}, the colours of a single blackbody at different
temperatures (solid line). In order to clarify why the YSOs have colours under
this curve we also plotted the colours of a blackbody modified by an emissivity
law $\propto\lambda^{-\beta}$ which better approximates the emission properties
of the dust, the main emitter in the FIR. The dashed line corresponds to the
case $\beta=2$, typical of interstellar dust, while the dotted line $\beta=1$,
is more representative of the dust in the star forming regions. It is clear
from the figure, that no positive $\beta$ can account for the positions of the
median values.

Another possible explanation is related to the confusion of the sources inside
the LWS beam. It is not possible to model this effect since we should know,
object by object, the physical properties and the spatial distribution of the
sources around the observed position. However, we can at least check if this
hypothesis moves the colours in the region of the plot where their medians lie,
by simply summing the fluxes of two blackbodies. In this case there are three
free parameters: the two temperatures and the ratio of their radii. As an
example, we have plotted in Figure~\ref{fig2}, the dotted-dashed line, the
colours obtained for the temperatures 25 and 90\,K, with a ratio varying from 0
to $\infty$ (the cross symbol marks the case corresponding to the radius of the
colder body 10 times larger than the radius of the warmer one). As can be seen
the sum moves the colours in the wanted direction.

The effects of a finite optical depth of the circumstellar envelope, while they
cannot be accounted for by a sum of blackbodies, should however be in
qualitative agreement with the previous model, so that we can expect also for
isolated objects a displacement from the blackbody colours as observed in our
plot.

\section{Conclusions}
The $[60-100]$ vs. $[100-170]$ two colour diagram has been presented for a
sample of YSOs observed with the LWS instrument on-board ISO. Each class of
YSOs occupies different regions of the diagram with bluer colours corresponding
to older classes. The median values of the colours are:\\
\begin{tabular}{rl}
&\\
Class 0:&$[60-100]=0.53\pm 0.23$\\
&$[100-170]=0.19\pm 0.18$\\
&\\
Class I:&$[60-100]=0.280\pm 0.075$\\
&$[100-170]=0.09\pm 0.10$\\
&\\
Bright Class I:&$[60-100]=0.02\pm 0.16$\\
&$[100-170]=-0.130\pm 0.083$\\
&\\
Class II:&$[60-100]=0.10\pm 0.16$\\
&$[100-170]=-0.04\pm 0.12$\\
&
\end{tabular}\\
The relative position of the different classes is related to the physical
properties of the circumstellar matter and is not an artefact of our data
analysis. This has been tested, for the [60-100] colour, through a comparison
with IRAS data which also show this segregation effect. It is not possible to
make such test for the other colour. However it is very unlikely that the class
separation is due to a bias either in the sample selection or in the data
analysis. The change of the colours is a genuine evolutionary effect: in the
younger objects we can only observe the outer, and colder, regions of the dusty
envelope. The median colours are systematically under the curve of the
blackbody colours, and the values we found are probably biased by the
combination of both the background contamination and the confusion of the
observed sources in the LWS beam. The low spatial resolution of the instrument
is indeed the main limitation of our result.

In this respect this work will be greatly improved by future space missions
which will observe with a higher spatial resolution. For instance, the multiband
photometric surveys of the FIR camera PACS at $\sim$60, 100 and 170$\,\mu$m,
will be able to make the census of each different evolutionary class of
protostellar and pre-main sequence objects in the most important star forming
molecular clouds.

In spite of the present limitation, we conclude that this two colour diagram
offers a powerful and simple tool for the evolutionary classification of YSOs.
The advantage over other methods is that follow up at other wavelengths to
identify the evolutionary status will not be needed. As an example, even if we
stress that with the LWS data our diagram works mainly as a statistical tool,
with only three fluxes, and no other photometric measurement, it is possible to
assign the evolutionary class to the source IRAS 18148$-$0440: Class~I, in
agreement with \citet{taf}, rather than Class~0 source \citep{AWB2}.

\section*{Acknowledgments}
This research has made use of the SIMBAD database, operated at CDS, Strasbourg,
France. The ISO Spectral Analysis Package (ISAP) is a joint development by the
LWS and SWS Instrument Teams and Data Centres. Contributing institutes are CESR,
IAS, IPAC, MPE, RAL and SRON. LIA is a joint developement of the ISO-LWS
Instrument Team at Rutherford Appleton Laboratories (RAL, UK - the PI Institute)
and the Infrared Processing and Analysis Center (IPAC/Caltech, USA).

We thank the anonymous referee for the valuable comments and remarks that
significantly improved the manuscript, and we acknowledge fruitful discussions
with C. Codella and S. Molinari.

\bsp 

\label{lastpage}

\end{document}